\documentclass[aps,pra,twocolumn,longbibliography,floatfix]{revtex4-2}

\newcommand{\sle}{\eta_\text{link}^\text{sys}}

\usepackage{graphicx,bm,natbib,upgreek,amsmath,mathrsfs,accents}
\usepackage{amsbsy}
\usepackage[dvipsnames]{xcolor}
\definecolor{myblue}{named}{MidnightBlue}
\definecolor{mygreen}{RGB}{0,120,0}
\usepackage[colorlinks=true,citecolor=myblue,linkcolor=myblue,urlcolor=myblue]{hyperref}
\usepackage[T1]{fontenc}
\usepackage{newtxtext,newtxmath}

\usepackage[scaled]{helvet}

\usepackage[caption=false]{subfig}
\usepackage{tabularx}

\usepackage[english]{isodate,babel}

\usepackage{amsthm}

\def\>{\rangle} 
\def\<{\langle}

\makeatletter
\def\thickhline{%
  \noalign{\ifnum0=`}\fi\hrule \@height \thickarrayrulewidth \futurelet
   \reserved@a\@xthickhline}
\def\@xthickhline{\ifx\reserved@a\thickhline
               \vskip\doublerulesep
               \vskip-\thickarrayrulewidth
             \fi
      \ifnum0=`{\fi}}
\makeatother

\newlength{\thickarrayrulewidth}
\DeclareGraphicsExtensions{.pdf, .jpg, .eps, .svg, .png}

\begin{document}

\title{Finite key effects in satellite quantum key distribution}

\author{Jasminder S. Sidhu}
\email{jasminder.sidhu@strath.ac.uk}
\thanks{These two authors contributed equally}
\affiliation{SUPA Department of Physics, University of Strathclyde, Glasgow, G4 0NG, United Kingdom}
\author{Thomas Brougham}
\email{t.brougham@strath.ac.uk}
\thanks{These two authors contributed equally}
\affiliation{SUPA Department of Physics, University of Strathclyde, Glasgow, G4 0NG, United Kingdom}
\author{Duncan McArthur}
\email{duncan.mcarthur@strath.ac.uk}
\affiliation{SUPA Department of Physics, University of Strathclyde, Glasgow, G4 0NG, United Kingdom}
\author{Roberto G. Pousa}
\email{roberto.gonzalez-pousa@strath.ac.uk}
\affiliation{SUPA Department of Physics, University of Strathclyde, Glasgow, G4 0NG, United Kingdom}
\author{Daniel K. L. Oi}
\email{daniel.oi@strath.ac.uk}
\thanks{Corresponding author}
\affiliation{SUPA Department of Physics, University of Strathclyde, Glasgow, G4 0NG, United Kingdom}

\date{\today}

\begin{abstract}
Global quantum communications will enable long-distance secure data transfer, networked distributed quantum information processing, and other entanglement-enabled technologies. Satellite quantum communication overcomes optical fibre range limitations, with the first realisations of satellite quantum key distribution (SatQKD) being rapidly developed. However, limited transmission times between satellite and ground station severely constrains the amount of secret key due to finite-block size effects. Here, we analyse these effects and the implications for system design and operation, utilising published results from the Micius satellite to construct an empirically-derived channel and system model for a trusted-node downlink employing efficient BB84 weak coherent pulse decoy states with optimised parameters. We quantify practical SatQKD performance limits and examine the effects of link efficiency, background light, source quality, and overpass geometries to estimate long-term key generation capacity. Our results may guide design and analysis of future missions, and establish performance benchmarks for both sources and detectors.
\end{abstract}

\maketitle

\section{Introduction}
\label{sec:introduction}

\noindent
Quantum technologies have the potential to enhance the capability of many applications~\cite{dowling2003quantum} such as sensing~\cite{Proctor2018_PRL, Sidhu2021_PRX, Sidhu2020_AVS}, communications~\cite{Scarani2009_RMP, Pirandola2020_AOP, Sidhu2020_PRXQ}, and computation~\cite{Knill2001_N}. Ultimately, a worldwide networked infrastructure of dedicated quantum technologies, i.e. a quantum internet~\cite{Wehner2018_S}, could enable distributed quantum sensors~\cite{Humphreys2013_PRL, Komar2014_N, Polino2019_O, Guo2020_NP}, precise timing and navigation~\cite{Okeke2018_NPJQI,Jozsa2000_PRL,Qian2020_arxiv}, and faster data processing through distributed quantum computing~\cite{Fitzsimons2017_NPJQI}. This will require the establishment of long distance quantum links at global scale. A fundamental difficulty is exponential loss in optical fibres which limits direct transmission of quantum photonic signals to ${<}1000$~km~\cite{Yin2016_PRL,Boaron2018_PRL,boaron2020progress,Chen2020_PRL}. Quantum repeaters may overcome the direct transmission limit but stringent performance requirements render them impractical by themselves for scaling to the intercontinental ranges needed for global scale-up~\cite{sidhu2021advances}. Alternatively, satellite-based free-space transmission significantly reduces the number of ground quantum repeaters required~\cite{Gundogan2020_arxiv}.

Satellite-based quantum communication has attracted much recent research effort~\cite{Hughes1999_SPIE,Kurtsiefer2002_N,Oi2017_EPJ,Mazzarella2020_C,polnik2020scheduling,villar2020entanglement,Gundogan2020_arxiv,sidhu2021advances} following recent in-orbit demonstrations of its feasibility by the Micius satellite~\cite{jianwei2018progress}. For low-Earth orbit (LEO) satellites, a particular challenge is the limited time window to establish and maintain a quantum channel with an optical ground station (OGS). For satellite quantum key distribution (SatQKD), this constrains the amount of secret key that can be generated due to two issues. First, the commonly assumed asymptotic resource assumption is not a good approximation for short received signal blocks. Without an arbitrarily large number of received signals, statistical uncertainties can no longer be ignored and the security of the distilled secret key requires careful treatment of the statistical fluctuations in estimated parameters~\cite{Tomamichel2012_NC,Lim2014_PRA,Rusca2018_APL}. Second, the trade-off between the proportion of signals used for parameter estimation and key generation becomes increasingly important to optimise. Further post-processing operations, such as error correction, reduce the amount of extractable secret key, with small block lengths leading to additional inefficiencies over the asymptotic limit~\cite{Tomamichel2017_QIP,Yin2020_SR}.

Initial SatQKD studies used the observed standard deviation to estimate statistical uncertainties and derive correction terms to the secret key rate~\cite{Bourgoin:2013fk,bourgoin2015experimental}. Analyses based on smooth entropies~\cite{Tomamichel2012_NC} improve finite-key bounds~\cite{Curty2014_NC} and have been applied to free-space quantum communication experiments~\cite{bacco2013experimental}. Recently, tight bounds~\cite{Yin2020_SR} and small block analyses~\cite{lim2021security} further improve key lengths for finite signals. Here, we provide a detailed analysis of SatQKD secret key generation which utilises tight finite block statistics in conjunction with system design and operational considerations.

As part of our modelling, we implement tight statistical analyses for parameter estimation and error correction to determine the optimised, finite-block, single-pass secret key length (SKL) for weak coherent pulse (WCP) efficient BB84 protocols using three signal intensities (two-decoy states). We base our nominal system model on recent experimental results reported by the Micius satellite~\cite{Yin2020_N} and use a simple scaling method to extrapolate performance to other SatQKD configurations. The effects of different system parameters are explored, such as varying system link efficiencies, protocol choice, background counts, source quality, and overpass geometries. We also provide a simple estimation method to determine the maximum expected long-term key volume at a particular OGS latitude. The improvement from combining data from multiple satellite passes is also examined. Our model and analysis may guide the design and specification of SatQKD systems, highlighting factors that limit secret key generation in the regime of high channel loss and limited pass duration.

We outline the system model in Section~\ref{sec:system_model}, including overpass geometries and system parameters. Operational models and finite block construction are given in Section~\ref{sec:operation_model}. We describe in Section~\ref{sec:finite_key_analysis} our SatQKD finite key analysis and examine the dependence of the SKL on different system parameters. We conclude and summarise our results in Section~\ref{subsec:future_work}.


\section{System Model}
\label{sec:system_model}

\begin{figure*}
    \centering
    \includegraphics[width=\linewidth]{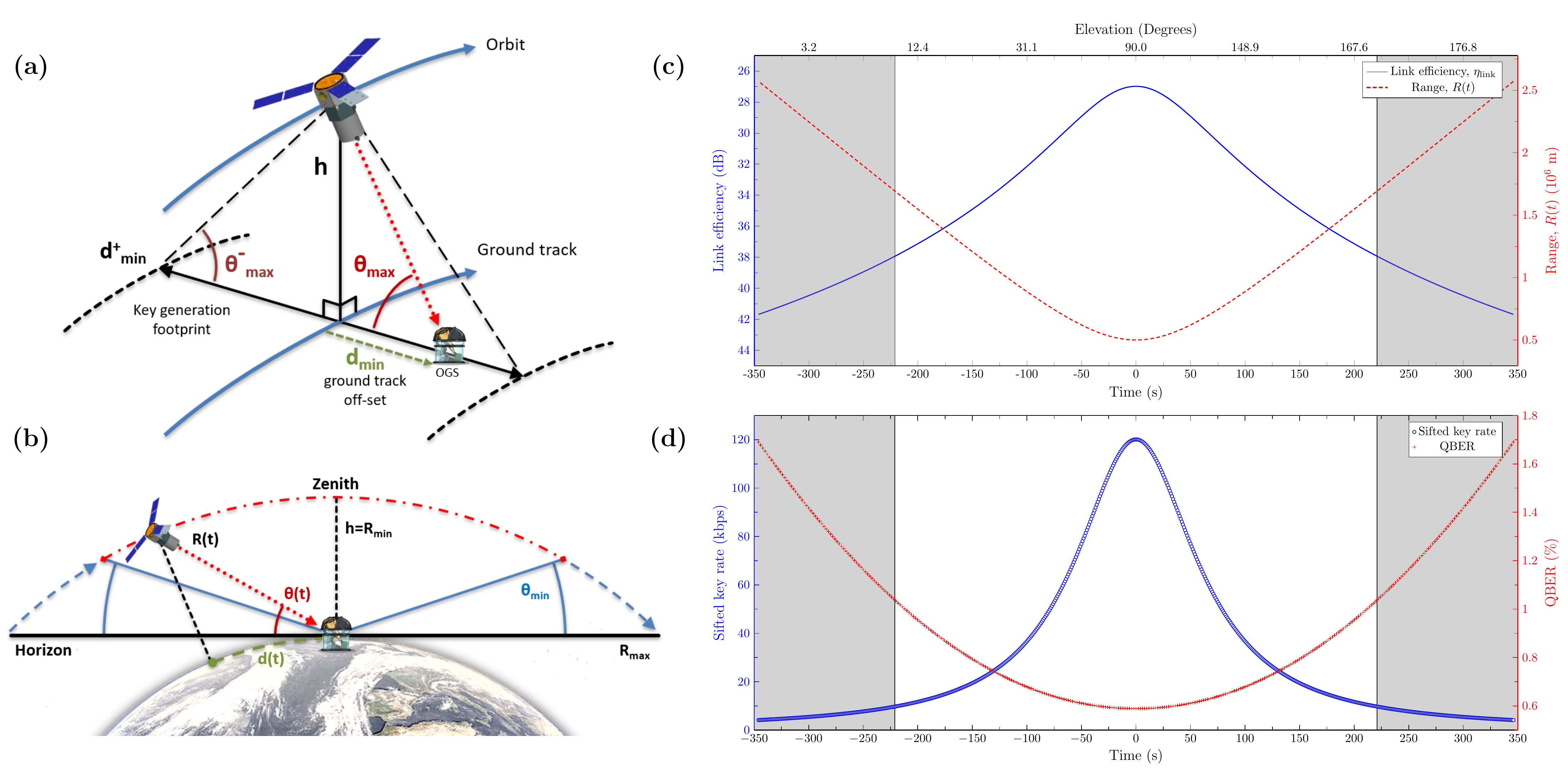}
    \caption{Satellite-to-ground QKD model. (a) General satellite overpass geometry for circular orbit of altitude $h$~\cite{vasylyev2019satellite}. Maximum elevation $\theta_\text{max}$ reached when satellite - OGS ground track distance is at a minimum, $d_\text{min}$. The smallest $\theta_\text{max}$, $\theta_\text{max}^{-}$, that generates finite key defines the key-generation footprint $2 d_\text{min}^+$. (b) Zenith overpass maximises transmission time with lowest average channel loss. Quantum transmission is limited to above $\theta_\text{min}$. (c) Zenith overpass channel loss vs time/elevation (based on Ref.~\cite{Yin2020_N} data). Link efficiency varies with range (\smash{$R^{-2}$} diffraction fall-off) and the atmospheric optical depth (attenuation and turbulence). Peak link efficiency $\smash{\eta_\text{link}}$ at zenith characterises the system link performance level, denoted $\smash{\sle}$. Additional constant losses (independent of range and elevation) are modelled by constant dB offset to the link efficiency curve, hence also off-setting $\sle$. (d) Modelled sifted key rate and QBER vs time/elevation. Zenith overpass, $\smash{\sle=27}$ dB, $\smash{p_\text{ec}=5\times 10^{-7}}$, and $\smash{\text{QBER}_\text{I}=0.5\%}$. BB84 WCP-DS protocol parameters: $f_s=200$ MHz, $\mu_{1,2,3}=0.5,0.08,0.0$, $p_{1,2,3}=0.72,0.18,0.1$, with $p_\mathsf{X}=0.889,0.9$ for the transmitter and receiver basis bias respectively, as in~\cite{Chen2021_N}. Shaded region indicates elevation below $\theta_\text{min} = 10^\circ$.}
    \label{fig:system_model}
\end{figure*}

\noindent
We consider a satellite in a circular Sun-synchronous orbit (SSO) of altitude $h=500$ km, similar to the Micius satellite~\cite{jianwei2018progress}, performing downlink QKD to an OGS at night to minimise background light. The elevation and range is calculated as a function of time for different overpass geometries representing different ground track offsets and maximum elevations for the overpass (Fig.~\ref{fig:system_model}(a)). The instantaneous link efficiency $\smash{\eta_\text{link}}$ is calculated as a function of the elevation $\theta(t)$ and range $R(t)$ to generate expected detector count statistics. The quantum link is restricted to be above $\theta_\text{min}=10^\circ$.

The link efficiency $\eta_\text{link}=-10\log_{10}{p_\text{d}}$ (dB) is determined by the probability, $p_\text{d}$, that a single photon transmitted by the satellite is detected by the OGS. A lower dB value of $\eta_\text{link}$ represents better system electro-optical efficiency, e.g. using larger transmit and receive aperture diameters, better pointing accuracy, lower receiver internal losses, and higher detector efficiencies. Internal transmitter losses are not included since they can be countered by adjusting the WCP source to maintain the desired exit aperture intensities~\cite{Bourgoin:2013fk}. We also do not consider explicitly time-varying transmittance, modelling the average change in channel loss due only to the change in elevation with time. For discrete variable QKD (DV-QKD) protocols, e.g. BB84, channel fluctuations do not directly impact the secret key rate, in contrast to continuous variable QKD where this appears as excess noise leading to key reduction~\cite{Usenko2012_NJP,hosseinidehaj2020composable}.

The ideal overpass corresponds to the satellite traversing the OGS zenith (Fig.~\ref{fig:system_model}(b)) giving the longest transmission time with lowest average channel loss. Generally, an overpass will not pass directly overhead but will reach a maximum elevation $\theta_\text{max}({<}90^\circ)$. We model total losses, including pointing and atmospheric effects, using Micius in-orbit measurements (Ref~\cite{Yin2020_N} Extended data, Fig.~3b) to construct a representative $\eta_\text{link}$ vs $\theta$ curve extrapolating over the entire horizon to horizon passage time (Fig.~\ref{fig:system_model}(c)). The Micius data represent a near ideal scenario since the OGSs are situated in dark sky conditions at high altitudes of ${\sim}3000$~m minimising the effects of atmospheric turbulence and attenuation, other sites may perform differently. Though the link efficiency in~\cite{Yin2020_N} is for the entanglement distribution system, not the optimised prepare-measure QKD downlink system as reported in~\cite{Chen2021_N}, it should still be representative of downlink efficiencies and sifted key rates achievable with current technologies (Fig.~\ref{fig:system_model}(d)). 

Systems with worse link performance are also considered. The system link efficiency $\sle$, characterising the overall system electro-optical efficiency independent of the overpass geometry, is defined as the $\eta_\text{link}$ value at zenith, i.e. the maximum probability of detecting a single photon sent from the satellite to the OGS. The baseline $\sle$ value is $27$ dB, corresponding to the improved Micius system using a $1.2$~m diameter OGS receiver at Delingha~\cite{Yin2020_N}. We scale $\eta_\text{link}$ uniformly and derive a worse (higher) $\sle$ value to model systems with greater fixed losses, e.g. using a smaller OGS, but otherwise similar behaviour to the Micius system. If the time-averaged ground spot size is much larger than the OGS diameter $D_r$, $\sle$ scales as $20\log_{10}(D_{r}/D_{r}^{0})$ (dB) where $D_{r}^{0}=1.2$ m is the reference Delingha OGS diameter.

Estimating the effect of transmitter aperture ($D_t$) is more complex since factors other than diffraction, such as pointing performance and turbulence, also determine the time-averaged ground spot size~\cite{liorni2019satellite,vasylyev2016atmospheric}. Micius, with $D_t$ of $180$~mm and $300$~mm and sub-$\mu$rad pointing performance, reported 10 $\mu$rad beam widths~\cite{Yin2020_N} which suggests the presence of non-diffraction-limited beam spreading effects that may result in a smaller dependence of $\sle$ on $D_t$. A smaller $D_t$ could result in a smaller increase in $\sle$ than given by purely diffractive beam broadening, conversely using a larger $D_t$ may not significantly improve $\sle$ if pointing and turbulence losses dominate. We refer the reader to detailed analyses of atmospheric turbulence~\cite{andrews2005laser} or extinction~\cite{Modtra_inproceedings}, and more recent works where their effects on SatQKD are considered~\cite{liorni2019satellite,vasylyev2019satellite,Pirandola2020SatQKD_arxiv}.

We include quantum bit errors arising from dark counts, background light, source quality, basis misalignment, and receiver measurement fidelity, simplifying our model by combining several quantities. The sum of dark and background light count rates, $p_\text{ec}$, is assumed constant and independent of elevation. All other errors are combined into a constant intrinsic quantum bit error rate $\text{QBER}_\text{I}$. Baseline system parameters are summarised in Table~\ref{tab:loss_error_parameters}.


\section{SatQKD Operations}
\label{sec:operation_model}

\noindent
A standard model of SatQKD uses large, fixed, and long-term OGSs to establish links with satellites~\cite{polnik2020scheduling}. A more demanding scenario is where the OGS may only be able to communicate sporadically with a particular satellite, limiting the amount of data that can be processed as a single large block, e.g. smaller, mobile OGS terminals may be required to generate a key from a limited number of passes, possibly only one, due to operational constraints. In contrast, fibre-based QKD can often assume a stable quantum channel able to be operated continuously until a sufficiently large block size is attained. High count rates and large block sizes, e.g. $10^{12}$, are generally more feasible in fibre-based QKD.

In SatQKD, often the theoretical instantaneous asymptotic key rate $\mathscr{R}_\infty(t)$ is integrated over the overpass to give the continuous secret key length (SKL)~\cite{polnik2020scheduling,Mazzarella2020_C},
\begin{align}
\label{eq:sklasymptotic}
    \text{SKL}_\infty^\text{Cont.}=\int_{t_\text{start}}^{t_\text{end}} \,\mathscr{R}_{\infty} (t) dt,
\end{align}
where the quantum transmission occurs between times $t_\text{start}$ and $t_\text{end}$. Data segments from multiple passes with identical statistics should be combined to yield an asymptotically large block for post-processing. More practically, small blocks, each having similar statistics, from different passes are combined to give the following SKL,
\begin{align}
    \text{SKL}_\infty^\text{Block}=\sum_j \mathscr{R}_\infty^{(j)} \mathcal{L}_j,
    \label{eq:asympblockrate}
\end{align}
where $\mathscr{R}_\infty^j$ is the asymptotic key rate for a small segment $j$, and $\mathcal{L}_j$ its length.
Operationally, this leads to considerable latency between establishing a first satellite-OGS link and the generation of secret keys after a sufficient number of subsequent overpasses. A less restrictive mode of operation is to combine data from multiple passes without segmenting into data blocks with similar statistics, and processing using asymptotically determined or assumed parameters. However, the security guarantee for this procedure requires closer examination.

\setlength{\thickarrayrulewidth}{2.1\arrayrulewidth}
\renewcommand{\arraystretch}{1.3}
\setlength{\tabcolsep}{8pt}
\begin{table}[b!]
  \centering
  \begin{tabular}{m{4.4cm}m{1.3cm}m{1.3cm}}
    \thickhline
    \textbf{Description}\vspace{2pt} & \textbf{Parameter}\vspace{2pt} & \textbf{Value} \vspace{2pt}\\
    \hline
    Intrinsic QBER & $\text{QBER}_\text{I}$ & $5\times10^{-3}$\\
    Afterpulse probability & $p_\text{ap}$ & $1\times 10^{-3}$\\
    Extraneous count probability/pulse & $p_\text{ec}$ & $5\times 10^{-7}$\\
    Source Rate & $f_s$ & $1\times 10^{8}$ Hz\\
    Correctness parameter & $\epsilon_\text{c}$ & $10^{-15}$\\
    Secrecy parameter & $\epsilon_\text{s}$ & $10^{-9}$ \\
    Error correction efficiency & $\lambda_\text{EC}$ & See text\\
    Baseline system link efficiency & $\sle$ & $27$ dB\\
    Orbital altitude & $h$ & $500$ km \\
    Minimum transmission elevation & $\theta_\text{min}$ & $10^\circ$ \\
    \thickhline
  \end{tabular}
  \caption{Baseline SatQKD system parameters. $\lambda_\text{EC}$ depends on block size and $\text{QBER}_\text{I}=0.5\%$ is consistent with Micius results~\cite{Yin2016_PRL}. We sum detector dark count and background rate in $p_\text{ec}$. A reported background count rate $500-2000$ cps per detector (Moon position dependent) lower bounds $p_\text{ec}$ by $5\times 10^{-7}$~\cite{yin2017satellite}, assuming a $1$ ns coincidence window.}
  \label{tab:loss_error_parameters}
\end{table}%

Here instead, we process overpass data as a single block without segmentation, incorporating finite statistics and uncertainties to maintain high levels of composable security with
\begin{align}
    \text{SKL}_\text{finite}=\text{SKL}\left(\left\{n_k^\mu,m_k^\mu\right\}\right),
\end{align}
where $\{n_k^\mu,m_k^\mu\}$ denote agglomerated observed counts without partitioning into sub-segments (see Methods~\ref{sec:finite_key_theory}). This is more practical for large constellations~\cite{vergoossen2020modelling,Mazzarella2020_C} and OGS numbers, obviating the need to track and store a combinatorially large number of link segments until each has attained a sufficiently large block size for asymptotic key extraction.


\section{Finite Key Length Analysis}
\label{sec:finite_key_analysis}

\noindent
We now quantify SatQKD system performance and SKL generation from different satellite overpasses. We employ the efficient BB84 protocol with weak coherent pulses (WCPs)~\cite{Pirandola2020_AOP} for which tight finite-key security bounds have been derived for one~\cite{Rusca2018_APL} and two~\cite{Lim2014_PRA} decoy states. The performance of one and two-decoy states is similar, however using two decoy states allows better vacuum yield estimation, useful in high loss operation. We optimise the two-decoy state protocol parameters and the amount of overpass data used in a block to
explore the dependence of the single-pass SKL on different variables and derive an expected long term key volume.

The efficient BB84 protocol~\cite{lo2005efficient} encodes signals in $\mathsf{X}$ and $\mathsf{Z}$ bases with unequal probabilities $p_\mathsf{X}$ and $1 -p_\mathsf{X}$ respectively. One basis is used exclusively for key generation and the other only for parameter estimation. We choose to use the error rate of the announced sifted $\mathsf{Z}$ basis to bound leaked information from the sifted $\mathsf{X}$ basis raw key. Biased basis choice improves the sifting ratio whilst retaining security. In the asymptotic regime, the sifting ratio tends to 1, versus 0.5 for symmetric basis choice (original BB84). This sifting ratio advantage  persists in the finite key regime (Section~\ref{sec:BB84_variants}) resulting in a longer raw sifted key that reduces parameter estimation uncertainties and provides more raw key to distil. 

For the two decoy-state WCP BB84 protocol, the sender randomly transmits one of three intensities $\mu_j$ for $j \in \{1,2,3\}$ with probabilities $p_j$. For the purposes of the security proof, we assume the intensities satisfy $\mu_1>\mu_2>\mu_3=0$. The finite block secret key length is then given by~\cite{Lim2014_PRA},
\begin{align}
    \ell  = \Big\lfloor s_{\textsf{X},0} & + s_{\textsf{X},1} (1 - h(\phi_\mathsf{X}))\nonumber\\
    & - \lambda_{\text{EC}} - 6 \log_2 \frac{21}{\epsilon_\text{s}} - \log_2 \frac{2}{\epsilon_\text{c}}\Big\rfloor,
\label{eqn:skl_lim_result}    
\end{align}
where $s_{\textsf{X},0}$, $s_{\textsf{X},1}$ and $\phi_\mathsf{X}$, are the $\mathsf{X}$-basis vacuum yield, single-photon yield and phase error rates respectively. In contrast to fibre-optic based systems, the size of the sifted $\mathsf{X}$-basis data block cannot easily be fixed for satellite-based QKD. Instead the number of pulses $N$ sent per pass is determined by the source repetition rate and the time available during a satellite overpass. In the asymptotic sample size limit (Eq.~\ref{eq:asympblockrate}), the $\mathsf{X}$ and $\mathsf{Z}$ basis data block sizes are straightforwardly determined by $N$, the pulse detection probability (itself a function of time), and the sifting ratio. However, finitely sized samples generate observed statistics that deviate from asymptotic expectations. Taking this into account can significantly reduce the SKL and we employ correction terms $\smash{\delta^{\pm}_{\mathsf{X}(\mathsf{Z}),k}}$ that relate the expected and observed statistics for bases $\mathsf{X}(\mathsf{Z})$ with a $k$-photon state, using the tight multiplicative Chernoff bound~\cite{Yin2020_SR} (see Methods~\ref{sec:finite_key_theory}).

Error syndrome information is publicly announced to perform error correction. The number of bits thus leaked is denoted $\lambda_\text{EC}$, and is accounted for during privacy amplification. In the finite key regime, this information leakage has a fundamental upper bound $\lambda_\text{EC} \le \log \vert\mathcal{M}\vert$, where $\mathcal{M}$ characterises the set of syndromes in the information reconciliation stage~\cite{Tomamichel2017_QIP}. We use an estimate of $\lambda_\text{EC}$ that varies with block size (Eq.~\ref{eq:lambdaec}). 

We characterise the reliability and security of the protocol by two parameters, $\epsilon_\text{c}$ and $\epsilon_\text{s}$. The protocol is $\varepsilon=\epsilon_\text{c}$ + $\epsilon_\text{s}$-secure if it is $\epsilon_\text{c}$-correct and $\epsilon_\text{s}$-secret~\cite{Lim2014_PRA,Renner2006_thesis}. For the numerical optimisation, we take $\epsilon_\text{c}=10^{-15}$ and $\epsilon_\text{s} = 10^{-9}$. Conditioned on passing the checks in the error-estimation and error-correction verification steps, an $\epsilon_\text{s}$-secret key of length $\ell$ can be generated. 


%
\begin{figure}[t!]
    \centering
    \includegraphics[width=0.98\columnwidth]{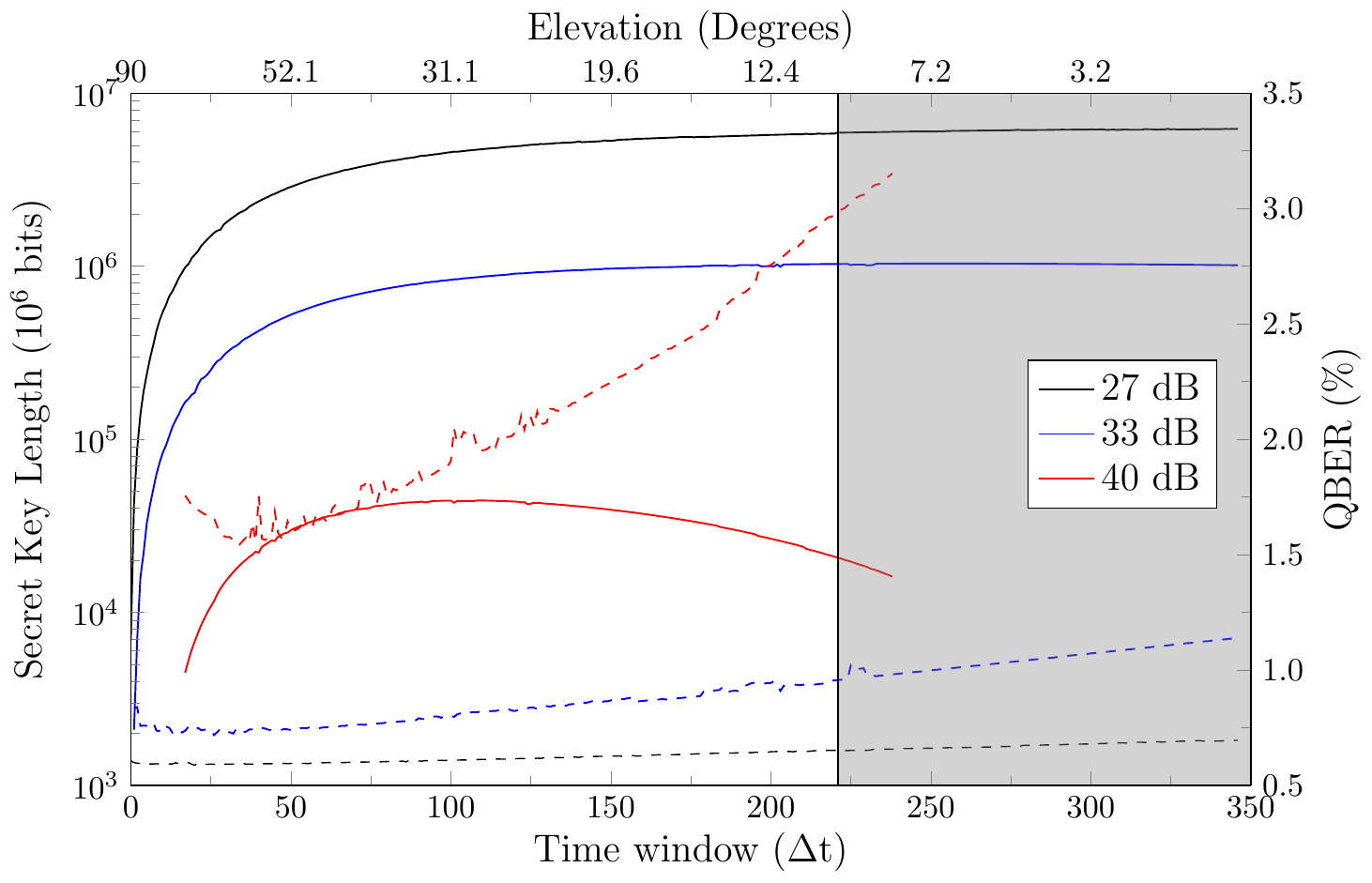}
    \caption{Secret key length and QBER vs transmission duration. Zenith overpass with different $\smash{\sle}$ (solid lines), $\smash{\text{QBER}_\text{I}=0.5\%}$, and $\smash{p_\text{ec}=5\times10^{-7}}$. Dashed lines represent truncated block QBERs. For each $\Delta t$, the SKL extractable from received data within $-\Delta t$ to $+\Delta t$ is optimised over protocol parameters. For the better $\smash{\sle}$ values, increasing $\Delta t$ beyond $200$ s leads to minor SKL improvement. For worse $\smash{\sle}$, including data from low elevations is detrimental to the SKL as seen by an increase in the truncated block averaged QBER. The non-smooth QBER appears since it is not the objective function of the optimisation. The shaded region indicates the time when the satellite elevation is lower than $\theta_\text{min} = 10^\circ$.}
    \label{fig:opt_time_window}
\end{figure}%

\subsection{Transmission time window optimisation}
\label{subsec:time_window}

\noindent
A satellite overpass is limited in duration and experiences highly varying channel loss, hence the expected count rates and QBER will change significantly throughout the pass. The received data obtained from lower elevations will have higher QBER compared to signals sent from higher elevations due to greater losses and the contribution of extraneous counts. This suggests that the SKL that could be extracted could be optimised by truncating poorer quality data from the beginning and end of the transmission period in some circumstances, despite resulting in a shorter raw block~\cite{wang2018prefixed}.

Our approach is to first fix the transmission duration and optimise the protocol parameters to use during the pass, then iterate over the window duration to find the highest resulting SKL. We define the transmission time window to run from $-\Delta t$ to $+\Delta t$, where $t=0$ represents the time of highest elevation $\theta_\text{max}$. For each $\Delta t$, we find optimum protocol parameters that maximise the SKL extractable from the data block generated within this transmission window. We impose a minimum elevation limit that reflects practicalities such as local horizon visibility and system pointing limitations. Here we use $\theta_\text{min}=10^\circ$ which for a zenith pass limits $2\Delta t$ to less than $\sim 440$ s.

We show the SKL as a function of $\Delta t$ for different $\sle$ values in Fig.~\ref{fig:opt_time_window}.  For good $\sle$, the QBER at low elevations does not rise greatly above $\text{QBER}_\text{I}$ and it is better to construct keys from the greatest amount of data where $\Delta t$ reaches the maximum allowed by $\theta_\text{min}$. Conversely, for poor $\sle$, utilising only data from near zenith leads to a longer SKL due to the better average QBER countering both the smaller raw key length and larger statistical uncertainties.

\begin{figure*}[t!]
    \centering
    \includegraphics[width=\linewidth]{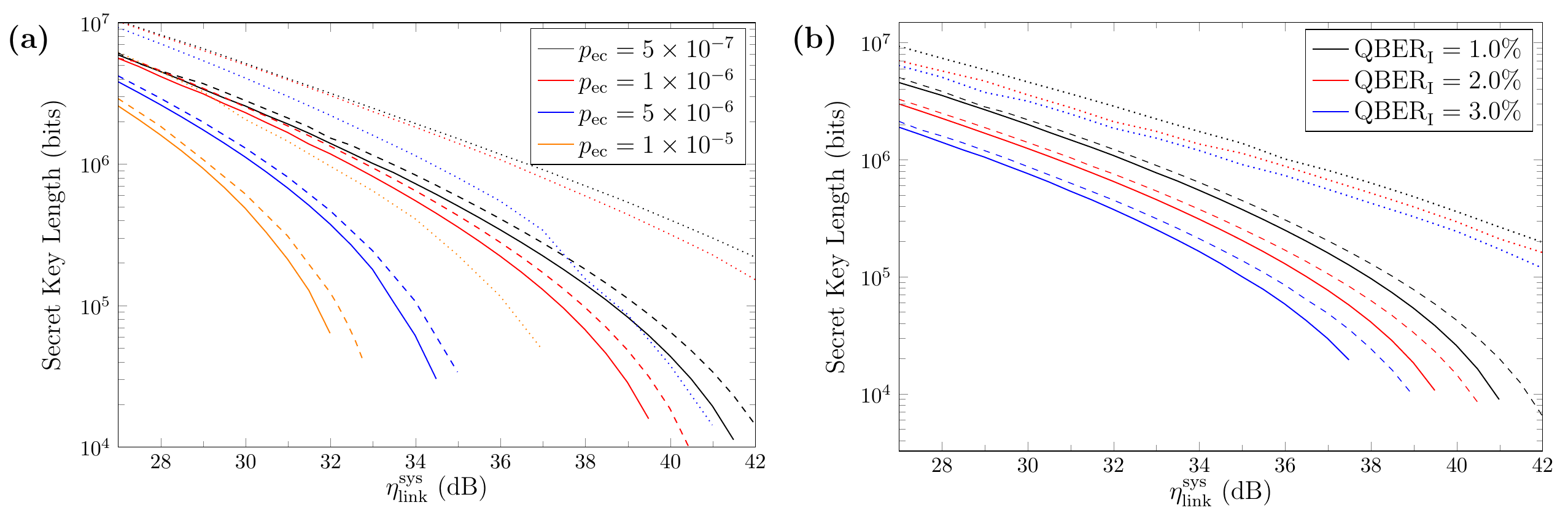}
    \caption{SKL with link efficiency. Fixed system parameters were $p_\text{ap}=10^{-3}$, $\theta_\text{min}=10^\circ$. In both plots, solid lines represent the optimised SKL with a single zenith overpass, dashed lines represent the two-pass normalised SKL, and dotted lines represent the normalised block asymptotic SKL (Methods~\ref{sec:asymptotic}). (a) SKL dependence on $p_\text{ec}$ with $\text{QBER}_\text{I} = 0.5\%$. The maximum $p_\text{ec}$ represents operation near a full Moon or with severe light pollution. (b) SKL dependence on $\text{QBER}_\text{I}$ with \smash{$p_\text{ec}= 5\times 10^{-7}$} per pulse.}
    \label{fig:SKL_dark_count_polarisation}
\end{figure*}%
%


\subsection{SKL system parameter dependence}
\label{subsec:sensitivity}

\noindent
We now determine the dependence of the SKL on different system parameters, including extraneous counts $p_\text{ec}$, intrinsic quantum bit errors $\text{QBER}_\text{I}$, and the source repetition rate $f_s$. The optimised SKL is then determined for different $\sle$ that model alternative SatQKD systems that differ from the baseline configuration (Table~\ref{tab:loss_error_parameters}) up to $\sle=42$ dB, corresponding to Micius transmitting to an OGS with $D_r=21.3$ cm, keeping all other system parameters the same.

We first consider how the SKL is affected by $p_\text{ec}$, which includes both detector dark counts and background light. For DV-QKD, silicon single photon avalanche photodiodes (Si-SPADs) are typically used for visible or near infrared wavelengths and can achieve a dark count rate of a few counts per second with thermoelectric cooling and temporal filtering~\cite{ceccarelli2020recent}. Superconducting nanowire single photon detectors (SNSPDs)~\cite{holzman2019superconducting} can offer superior wavelength sensitivity (particularly beyond 1 $\mu$m), dark count rate (less than 1 cps), and timing jitter, though at the expense of greater cost, size, weight, and power (SWaP), owing to the need for cryogenic operation. Background light, due to light pollution and celestial bodies (most notably the Moon) is the main constraint to minimising $p_\text{ec}$~\cite{er2005background}. We have used a simplified model which does not include elevation dependent background light levels (which is highly site dependent). The impact of varying $p_\text{ec}$ on the SKL is shown in Fig.~\ref{fig:SKL_dark_count_polarisation}(a). While extraneous counts increase the vacuum yield $s_{\mathsf{X},0}$, any addition to the SKL is offset by reductions from worse phase error rates and error correction terms. For example, a factor of 10 increase of $p_\text{ec}$ results in a 40\% net reduction to the SKL for $\smash{\sle}=27$ dB. The effect of extraneous counts is further compounded for worse $\sle$ values and can result in zero SKL due to an excessive QBER.

\begin{figure}[t!]
    \centering
    \includegraphics[width=\columnwidth]{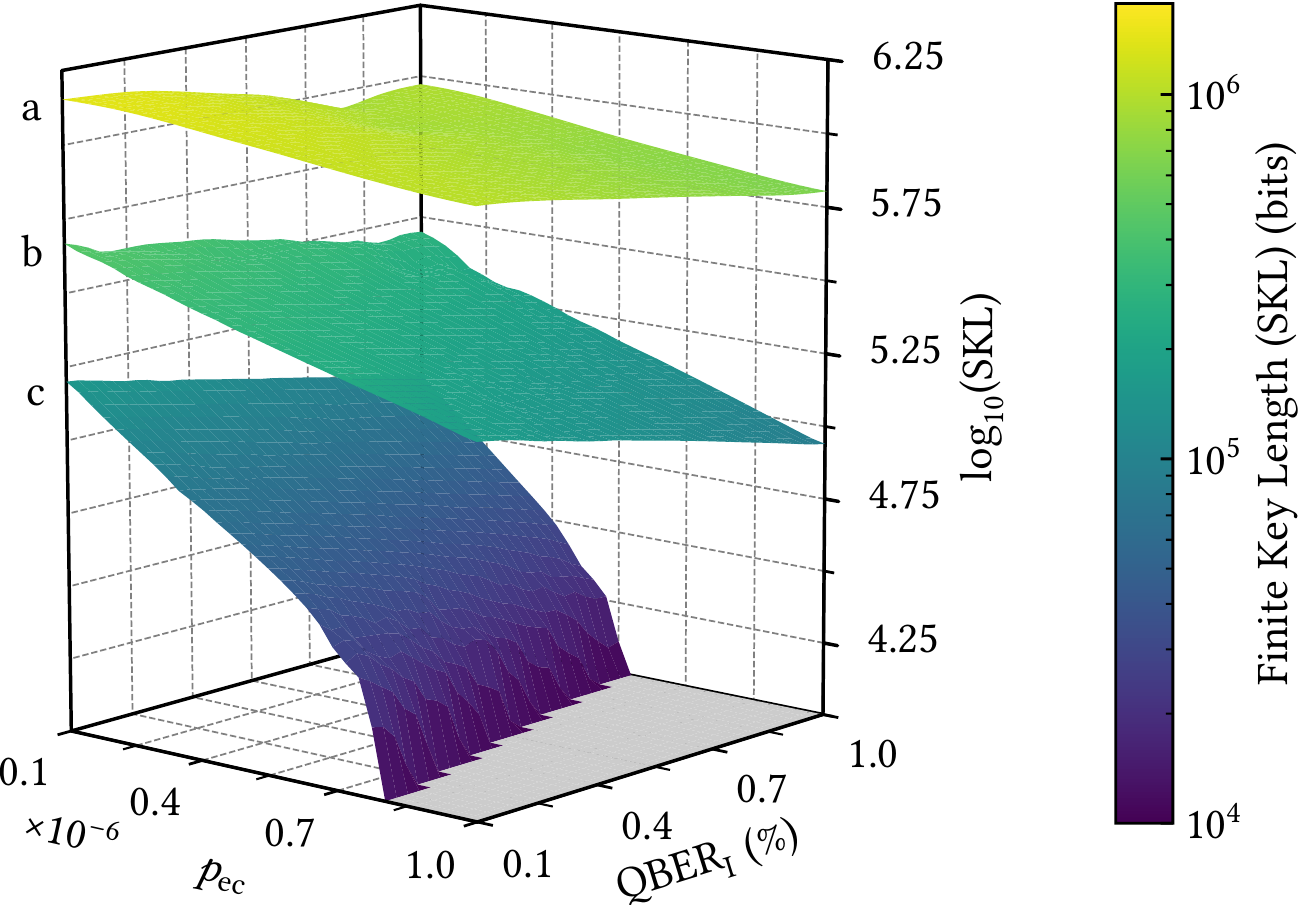}
    \caption{SKL versus $p_\text{ec}$ and $\text{QBER}_\text{I}$. Single zenith overpass with $\sle$ values: a) $33$ dB, b) $37$ dB, and c) $40.5$ dB. The grey region indicates zero SKL.}
    \label{fig:two_contour_plot}
\end{figure}

The QBER also suffers from effects such as non-ideal signals, satellite-OGS reference frame misalignment, or imperfect projective measurements by the OGS. We characterise these by an intrinsic system error, $\text{QBER}_\text{I}$, which is independent of the count rate or channel loss. Fig.~\ref{fig:SKL_dark_count_polarisation}(b) illustrates the effect of different $\text{QBER}_\text{I}$ on the SKL. We observe that the finite key length is not as susceptible to changes in the $\text{QBER}_\text{I}$ as compared with $p_\text{ec}$. The relative effects of both $p_\text{ec}$ and $\text{QBER}_\text{I}$ on the SKL is illustrated in Fig.~\ref{fig:two_contour_plot} (see also Extended Data Fig.~\ref{fig:contour_plot}). The SKL varies greatly along the $p_\text{ec}$ direction, with zero finite key returned for large $\smash{p_\text{ec}}$ irrespective of improvements to QBER$_\text{I}$. This indicates that improvements to background light suppression and detector dark count over source fidelities and satellite alignment should be prioritised.

We can estimate the effect of increasing the source rate $f_s$ by incorporating a correction factor to $\sle$ for the current results. Since the SKL is a function of $\{n_k^\mu,m_k^\mu\}$, these only depend on the integrated product of the source rate and the link efficiency, with all other system parameters kept the same. Therefore, a $100$ MHz source at a given $\sle$ provides the same amount of raw key as a $1$ GHz source with a $10$ dB worse system link efficiency, e.g. corresponding to a three times smaller OGS receiver diameter (Extended Data Fig.~\ref{fig:npulse}). This approximation, however, neglects extraneous counts and the resultant instantaneous QBER which is unaffected by the source rate but does depend on $\sle$. Nevertheless, the above heuristic holds provided the contribution of $p_\text{ec}$ to QBER is small and thus SKL is mostly constrained by raw key length and statistical uncertainties.

Practically, the amount of raw key that could be transmitted during an overpass may be limited by the amount of available stored random bits, irrespective of increases in $f_s$. Real-time quantum random number generation can overcome this though at the expense of increased SWaP which is often constrained on smaller satellites. Even with limits on the amount of raw key available, increasing $f_s$ can still be advantageous by compressing the transmission into a smaller duration around $\theta_\text{max}$, which decreases the average loss per transmitted block.

Protocol parameters that affect the SKL include signal intensities $\mu_j$, their probabilities $p_j$, and basis bias $p_\mathsf{X}$. The optimum values of these system parameters generally depend on $\sle$ and the overpass geometry. The optimal key generation basis choice probability $p_\mathsf{X}$ decreases with increasing $\sle$. As the OGS detects fewer photons with increasing loss, leading to worse parameter estimation of $s_{\mathsf{X},0}$, $s_{\mathsf{X},1}$ and $\phi_{\mathsf{X}}$ due to greater statistical fluctuations, to compensate we need to collect more $\mathsf{Z}$ basis events by increasing $1-p_\mathsf{X}$. The reduced number of key generation events is outweighed by better bounds on the key length parameters. This implies that the uncertainties in the parameter estimation from finite statistics dominates the SKL compared with raw key length when $\sle$ becomes poor (Extended Data Fig.~\ref{fig:optimisation_parameters}).

\subsection{SKL and overpass geometry}
\label{subsec:nonzenith}

\begin{figure}[t!]
    \centering
    \includegraphics[width=0.48\textwidth]{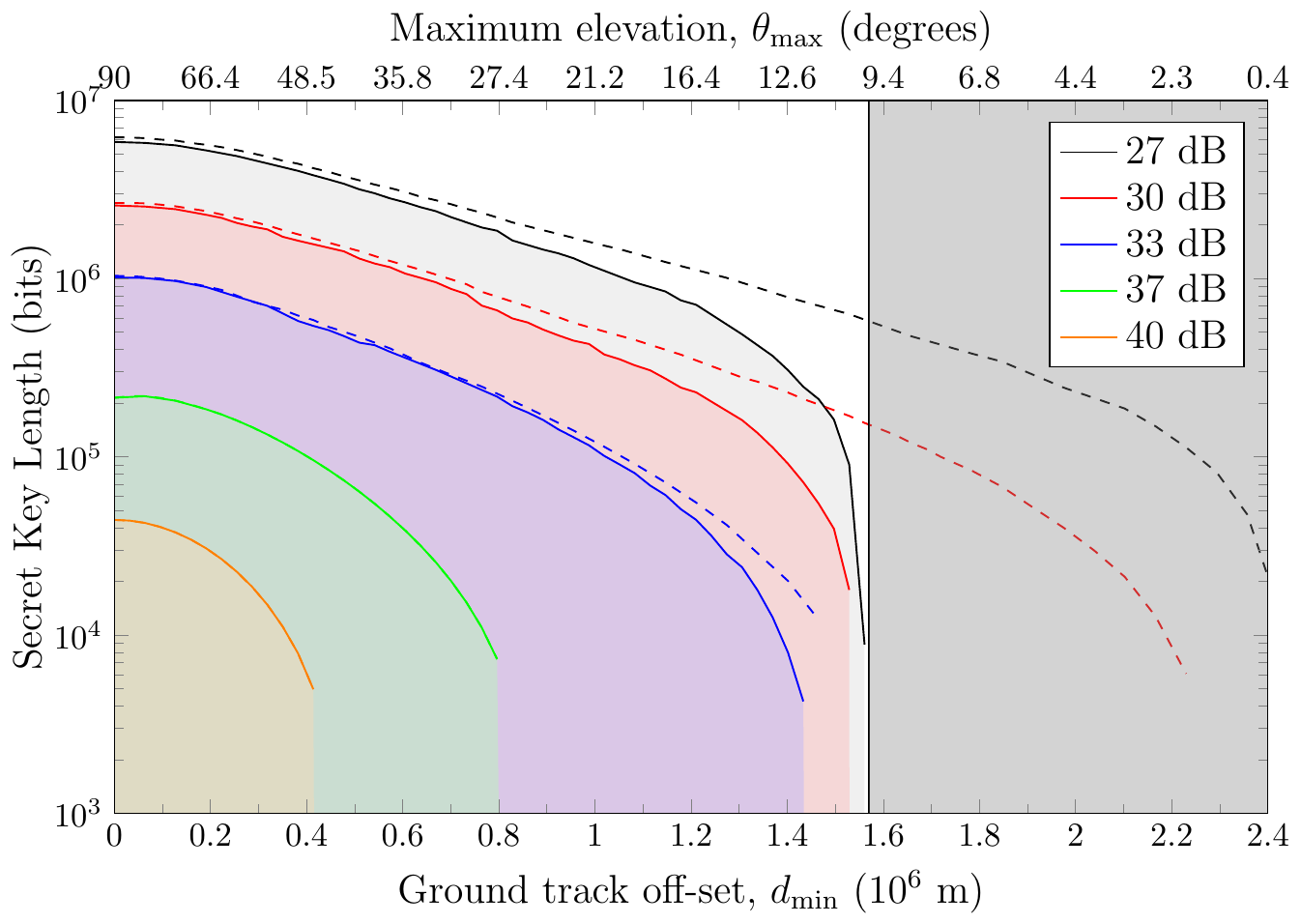}
    \caption{SKL vs ground track off-set. \smash{$p_\text{ec}=5\times 10^{-7}$} and $\text{QBER}_\text{I} = 0.5\%$. The key generation footprint is given by the maximum $d_\text{min}$ with non-zero SKL. Solid lines correspond to imposing $\theta_\text{min}=10^\circ$ (indicated by dark grey region on the right) and dashed lines to no elevation limit. The shaded areas under the curves determine the expected annual SKL for different $\sle$. Imposing $\theta_\text{min}=10^\circ$ reduces the area, and hence the expected annual SKL, by 18.82\%, 13.37\%, and 3.58\% at 27 dB, 30 dB, and 33 dB respectively.}
    \label{fig:key_elevation_dependence}
\end{figure}

\noindent
A typical satellite overpass will not go directly over zenith but will pass within some minimum ground track offset $d_\text{min}$ of the OGS, reaching a maximum elevation $\theta_\text{max} (< 90^\circ)$ (Fig.~\ref{fig:system_model}(a)). To maximise the number of overpass opportunities that can generate a secret key, a SatQKD system should be able to operate with as low a maximum elevation $\theta_\text{max}$ as possible. The SKL per pass as a function of $d_\text{min}$ is shown in Fig.~\ref{fig:key_elevation_dependence} for different $\sle$ values. As expected, overpasses with smaller $\theta_\text{max}$ deliver smaller SKLs due to shorter transmission times and lower count rates from worse average $\eta_\text{link}$ at lower elevations and longer ranges. The SKL vanishes once $\theta_\text{max}$ is below a critical elevation angle $\theta_\text{max}^-$ when the small block size leads to excessive statistical uncertainties or the average QBER becomes too high.

We now can estimate the long-term average amount of secret key that can be generated using single overpass blocks with an OGS site situated at a particular latitude. We first integrate the area under the SKL vs $d_\text{min}$ curve,
\begin{align}
    \text{SKL}_\text{int}=2\int_{0}^{d_\text{min}^+} \text{SKL}_{d_\text{min}} \; d d_\text{min},
\end{align}
where $d_\text{min}^+$ is the maximum OGS ground track offset that generates key. Assuming a sun synchronous orbit, we then estimate the expected annual key from (neglecting weather),
\begin{align}
    \overline{\text{SKL}}_\text{year}=N_\text{orbits}^\text{year}\frac{\text{SKL}_\text{int}}{L_\text{lat}},
\end{align}
where $N_\text{orbits}^\text{year}$ is the number of orbits per year, and $L_\text{lat}$ is the longitudinal circumference along the line of latitude at the OGS location. This estimate assumes that $d_\text{min}$ is evenly distributed (unless in an Earth synchronous orbit~\cite{Mazzarella2020_C}) and the OGS is not close to the poles where the orbital inclination (${\sim}97^\circ$) invalidates the approximation of even distribution. In Table~\ref{tab:SKL_year}, we summarise the expected yearly secret finite key lengths attainable for various $\sle$ at latitude $55.9^\circ$ N.

\setlength{\thickarrayrulewidth}{2.1\arrayrulewidth}
\renewcommand{\arraystretch}{1.4}
\setlength{\tabcolsep}{8pt}
\begin{table}[t!]
  \centering
  \begin{tabularx}{0.72\columnwidth}{llX}
    \thickhline
    \textbf{$\sle$ \quad}\vspace{2pt} & $\textbf{SKL}_{\textbf{int}}$\vspace{2pt} & $\overline{\textbf{SKL}}_{\textbf{year}}^{\textbf{55.9}^\circ \textbf{N}}\;$ \vspace{2pt}\\
    \hline
    27 dB & $3.74 \times 10^{12}$~bm & $0.9131$ Gb \\
    30 dB & $1.52 \times 10^{12}$~bm & $0.3720$ Gb \\
    33 dB & $5.40 \times 10^{11}$~bm & $0.1318$ Gb \\
    37 dB & $8.75 \times 10^{10}$~bm & $0.0214$ Gb \\
    40 dB & $1.13 \times 10^{10}$~bm & $0.0028$ Gb \\
    \thickhline
  \end{tabularx}
  \caption{Expected annual SKL for different $\sle$.  The $\text{SKL}_{\text{int}}$ values correspond to the area under each SKL vs $d_\text{min}$ curve in Fig.~\ref{fig:key_elevation_dependence} with units of bit-metres (bm). For $h=500$ km, $N_\text{orbits}^\text{year}{\sim}5500$, and at $55.9^\circ$ N (latitude of Glasgow) $L_{55.9^\circ N}\sim 2.25\times 10^7$ m, the expected annual key volume $\overline{\text{SKL}}_\text{year}=2.44\times 10^{-4} \text{SKL}_\text{int}\ \text{m}^{-1}$. We assume $\theta_\text{min}=10^\circ$, \smash{$p_\text{ec}=5\times 10^{-7}$}, and $\text{QBER}_\text{I} = 0.5\%$.}
  \label{tab:SKL_year}
\end{table}%
%


\subsection{Multiple satellite passes}
\label{subsec:results}

\noindent
Disregarding latency of key generation, data from several overpasses can be combined to improve SKL generation. Fig.~\ref{fig:SKL_dark_count_polarisation} displays finite and asymptotic SKLs for a single zenith overpass. The overpass-normalised asymptotic SKL corresponds to multiple overpasses where the block sizes used for key rate determination tend to infinity. Instead of Eq.~\ref{eq:asympblockrate}, we aggregate data from separate overpasses (with the same geometry and protocol parameters) into a single processing block without segmented instantaneous asymptotic key rates, only assuming asymptotically ascertained combined block parameters. The process of taking the block size to infinity requires care due to the limited amount of data per pass (Methods~\ref{sec:asymptotic}). The per-pass SKL increases significantly as the block asymptotic regime is approached.

\begin{figure}[t!]
    \centering
    \includegraphics[width=0.48\textwidth]{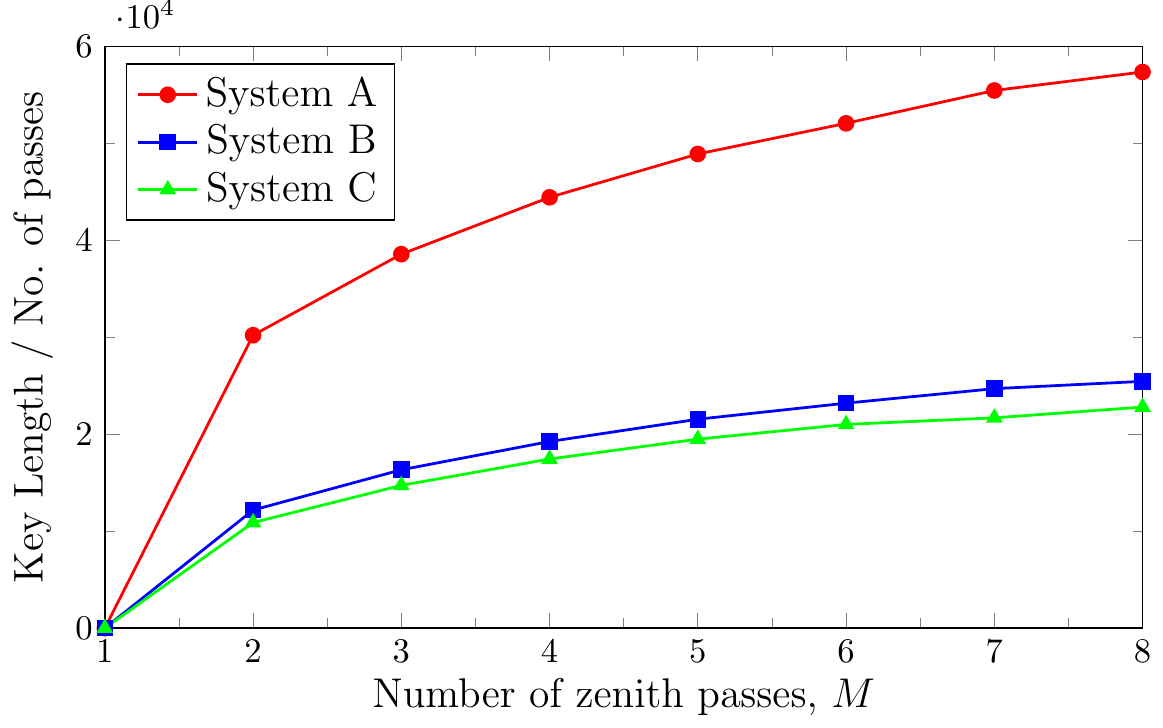}
    \caption{Per-pass SKL vs Number of combined overpasses. Secret key is generated from combined multiple overpass data. Zenith overpasses with $\theta_\text{min}=10^\circ$.  System parameter sets $\{\sle,p_\text{ec}, \text{QBER}_\text{I}\}$: A=$\{45.7\,\text{dB},10^{-7}, 0.5\%\}$, B=$\{44.8\,\text{dB},10^{-7}, 1\%\}$, and C=$\{40.5\,\text{dB},5\times10^{-7},1\%\}$.}
    \label{fig:SKL_multiple_passes}
\end{figure}

Systems with zero single-overpass SKL can generate positive key by accumulating signals from several overpasses (Fig.~\ref{fig:SKL_multiple_passes}). If $\ell_M$ is the total SKL generated from $M$ identical satellite passes, then $\smash{\ell_M \ge M \ell_1}$ with diminishing improvement $\ell_{M+1} - \ell_{M}$ with increasing $M$, with the largest jump going from $M=1$ to $2$. Averaging over several identical passes does not improve the underlying block averaged signal parameters such as QBER, hence the per-pass SKL improvement is mainly due to smaller estimation uncertainties from increased sample size, with better error correction efficiency and reduced $\lambda_\text{EC}$ also contributing. This shows that finite statistical fluctuations are the principle limitation to the SKL that can be generated with a limited number of overpasses.

Practically, employing multiple passes to improve SKL generation should be balanced against greater latency and potential security vulnerabilities in storing large amounts of raw data for a longer period between passes. This will depend on the assumed security model and anticipated attack surfaces of the encryption keys at rest~\cite{alleaume2018implementation}.


\subsection{Protocol performance}
\label{sec:BB84_variants}

\noindent
The choice of QKD protocol can also significantly affect the SKL due to finite block size statistics. This is illustrated by comparing two BB84 variants: efficient BB84 (considered thus far)~\cite{lo2005efficient} and standard BB84~\cite{Bennett1984_original}. The main difference is the basis choice bias, with standard BB84 choosing both $\mathsf{X}$ and $\mathsf{Z}$ bases with equal (symmetric) probability, while efficient BB84 allows biased (asymmetric) basis choice. Standard BB84 also uses both bases to generate key, hence requires parameter estimation of both. Efficient BB84 uses only one basis for key generation and the other for parameter estimation. We refer to the two protocols as s-BB84 and a-BB84 respectively.

\begin{figure}[t!]
    \centering
    \includegraphics[width=\columnwidth]{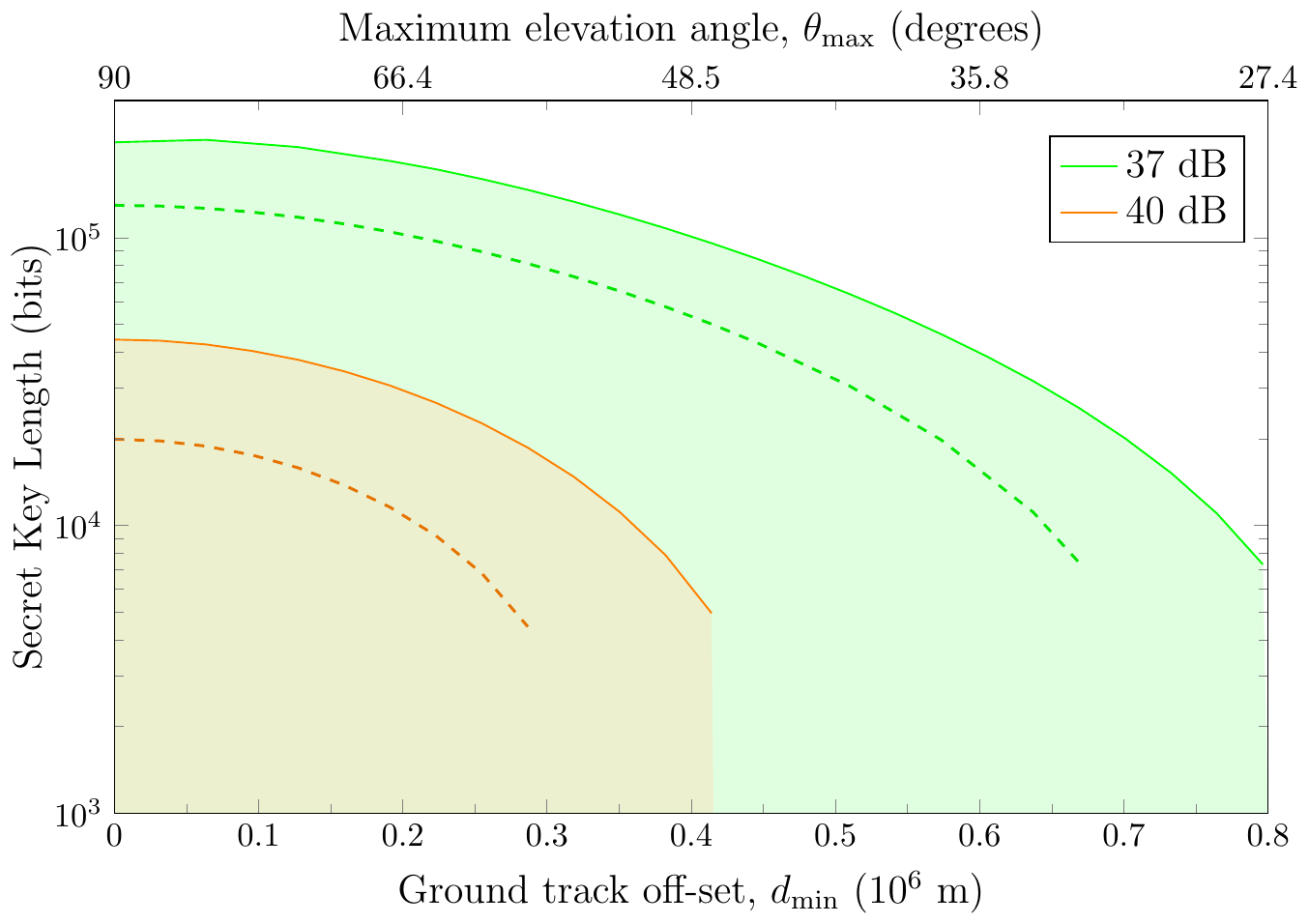}
    \caption{Protocol effect on SKL vs ground track offset. Shown are a-BB84 (solid lines) and s-BB84 (dashed lines) with $p_\text{ec}=5\times 10^{-7}$, $\text{QBER}_\text{I} = 0.5\%$.}
    \label{fig:asym_sym_elevation}
\end{figure}%

In general a-BB84 produces higher SKL for all system link efficiencies, can tolerate worse $\sle$, and operate with lower $\theta_\text{max}$ (Extended data Fig.~\ref{fig:asym_sym_loss}). The expected annual SKL (as in Sec.~\ref{subsec:nonzenith}) with a-BB84 and s-BB84 for the baseline system parameters (Table~\ref{tab:loss_error_parameters}) is obtainable from Fig.~\ref{fig:asym_sym_elevation}. For $\sle=37$ dB and $40$ dB, a-BB84 generates $86\%$ and $182\%$ times more key on average, respectively. The advantage increases with worse $\sle$ due to several reasons. First, the better sifting ratio of a-BB84 results in more raw bits that also allows for better parameter estimation. Second, a-BB84 uses all events in one basis to estimate the vacuum and single photon yields in the other. In contrast, s-BB84 reveals an optimally sized random sample of results for each basis to separately estimate signal parameters. Hence only half the revealed results are used for each basis estimation, leading to worse statistical uncertainties.

A tangential advantage is that a-BB84 requires less classical communication than s-BB84. In a-BB84 the choice of bits for public comparison for QBER estimation is implicit in the basis choice and automatically revealed during sifting. A minor disadvantage to a-BB84 is the extra overhead in generating biased basis probabilities~\cite{gryszka2020biased}, but the advantage of s-BB84 in this regard is relatively small taking into account all the biased probabilities required for both decoy-state variants.


\section{Conclusions and discussions}
\label{subsec:future_work}

\noindent
Important differences with fibre-based QKD mean that small sample statistical uncertainties have a significant impact on the performance of satellite QKD. The restricted overpass time of a LEO satellite constrains the amount of sifted key that can be established with an optical ground station. Operationally, a secret key may need to be generated from a single pass, thus statistical uncertainties will significantly impact performance in practical scenarios. Our study examines the severity of finite-key effects for representative space-ground quantum channel link efficiencies as indicated by the in-orbit demonstration of Micius.

Our results highlight the influence of system and protocol parameters on the secret key length that can be generated from a single overpass. For the range of system link efficiencies considered ($\sle=27\ \text{to} \ 42$ dB), the strongest dependence comes from the degree by which extraneous counts ($p_\text{ec}$) can be suppressed, with a much weaker dependence on the intrinsic signal/measurement quality ($\text{QBER}_\text{I}$) of the system. There is also a minimal effect of imposing a minimum elevation limit for quantum signal transmission. This suggests that SatQKD systems should prioritise background light suppression over higher intrinsic quantum signal visibilities or extending transmission closer to the horizon.

The dominance of finite-key effects is highlighted by the comparison between efficient (asymmetric basis bias) BB84 with conventional (symmetric basis bias) BB84. The much greater secret key length of efficient BB84 stems from a better sifting ratio and longer raw key length as well as obviating the need to perform parameter estimation for 2 bases that would further compound the finite statistical uncertainty. This greater performance translates into a higher secret key length for a given overpass geometry and also extends the satellite ground footprint within which secret key can be established with a ground station. Overall, improvements in system performance, whether through better protocols, system link efficiencies, or higher source rates, can significantly expand expected annual secret key volumes, e.g. a system link efficiency improvement of 3 dB from 40 dB to 37 dB improves the expected annual key volume by a factor of 7.6. Should operations allow, secret key extraction efficiency can also be enhanced by combining data blocks from several passes, especially if no secret is possible from a single overpass.

This preliminary study of SatQKD finite-key effects can be extended to remove some simplifications and approximations. An immediate extension would include a more comprehensive time and elevation dependent quantum channel model incorporating scattering, turbulence, and anisotropic background light distributions. Site dependent scenarios could include local horizon limits, light pollution, and seasonal weather effects. We can constrain the optimisation of the protocol parameters to reflect additional restrictions on system operations and deployment in practice. Ultimately, design and optimisation of SatQKD systems should incorporate orbital modelling of constellations and ground stations geographic diversity together with cost/performance trade-off studies.


\section*{Acknowledgements}

\noindent
We acknowledge support from the UK NQTP and the Quantum Technology Hub in Quantum Communications (EPSRC Grant Ref: EP/T001011/1), the UK Space Agency (NSTP3-FT-063, NSTP3-FT2-065, NSIP ROKS Payload Flight Model), the Innovate UK project ReFQ (Project number: 78161), and QTSPACE (COST CA15220). DO is an EPSRC Researchers in Residence at the Satellite Applications Catapult (EPSRC Grant Ref: EP/T517288/1). DO and TB acknowledge support from the Innovate UK project AirQKD (Project number: 45364). DO and DM acknowledge support from the Innovate UK project ViSatQT (Project number: 43037). RP acknowledges support from the EPSRC Research Excellence Award (REA) Studentship. The authors thank J. Rarity, D. Lowndes, S. K. Joshi, E. Hastings, P. Zhang, and L. Mazzarella. for insightful discussions. DO also acknowledges discussion with S. Mohapatra, Craft Prospect Ltd., and support from the EPSRC Impact Acceleration Account.


\section*{Author contributions}

\noindent
DO conceived, obtained funding for, and initiated the research. JSS and TB developed the finite key optimiser and performed the simulations with additional assistance from DMcA. RGP conducted background literature reviews. JSS, TB, and DO wrote an initial draft with inputs and feedback from all authors. All authors contributed equally in selecting relevant literature, final editing, and proofreading of the manuscript.

\section{Methods}

\subsection{Finite key analysis for decoy-state BB84}
\label{sec:finite_key_theory}

\noindent
The Bennett-Brassard 1984 (BB84) quantum key distribution (QKD) protocol is widely implemented owing to its simplicity, overall performance and provable security~\cite{Bennett1984_original}. However, practical implementations of BB84 depart from the use of idealised single-photon sources. Instead, weak pulsed laser sources are used given their wide availability and relative ease of implementation. This improves repetition rates over current single photon sources, but leaves the BB84 protocol vulnerable to photon-number-splitting (PNS) attacks that exploit the multi-photon pulse fraction present in emitted laser pulses~\cite{Brassard2000_PRL}.

Decoy-state protocols circumvent PNS attacks and improve tolerance to high channel losses, with minimal modification to BB84 implementations. These protocols employ multiple phase randomised coherent states with differing intensities that replace signal pulses. This modification permits better characterisation of the photon number distribution of transmitted pulses associated with detection events~\cite{Hwang2003_PRL}, which reliably detect the presence of PNS attacks in the quantum channel. Decoy-state BB84 protocols also allow better estimation of the secure fraction of the sifted raw key (vacuum and single photon yields), which makes them a secure and practical implementation of QKD.

The security of decoy-state QKD was initially developed assuming the asymptotic-key regime~\cite{Wang2005_PRL,Lo2005_PRL}. For applications with finite statistics, uncertainties in the channel parameters cannot be ignored~\cite{Ma2005_PRA, Hasegawa2007_arxiv, Cai2009_NJP}. Early approaches in handling these finite key statistics used Gaussian assumptions to bound the difference between the asymptotic and finite results~\cite{Zhang2017_PRA}. This restricts the security to collective and coherent attacks. Security analyses for more general attacks have also been developed~\cite{Hayashi2014_NJP}. The multiplicative Chernoff bound~\cite{Curty2014_NC,Zhang2017_PRA} and Hoeffding's inequality~\cite{Lim2014_PRA} can be used to bound the fluctuations between the observed values and the true expectation value. Recently, a more complete finite-key analysis for decoy-state based BB84, with composable security, has been presented in Ref.~\cite{Yin2020_SR}, which uses the multiplicative Chernoff bound to derive simple analytic expressions that are tight.

\begin{figure}[t!]
    \centering
    \includegraphics[width=0.98\columnwidth]{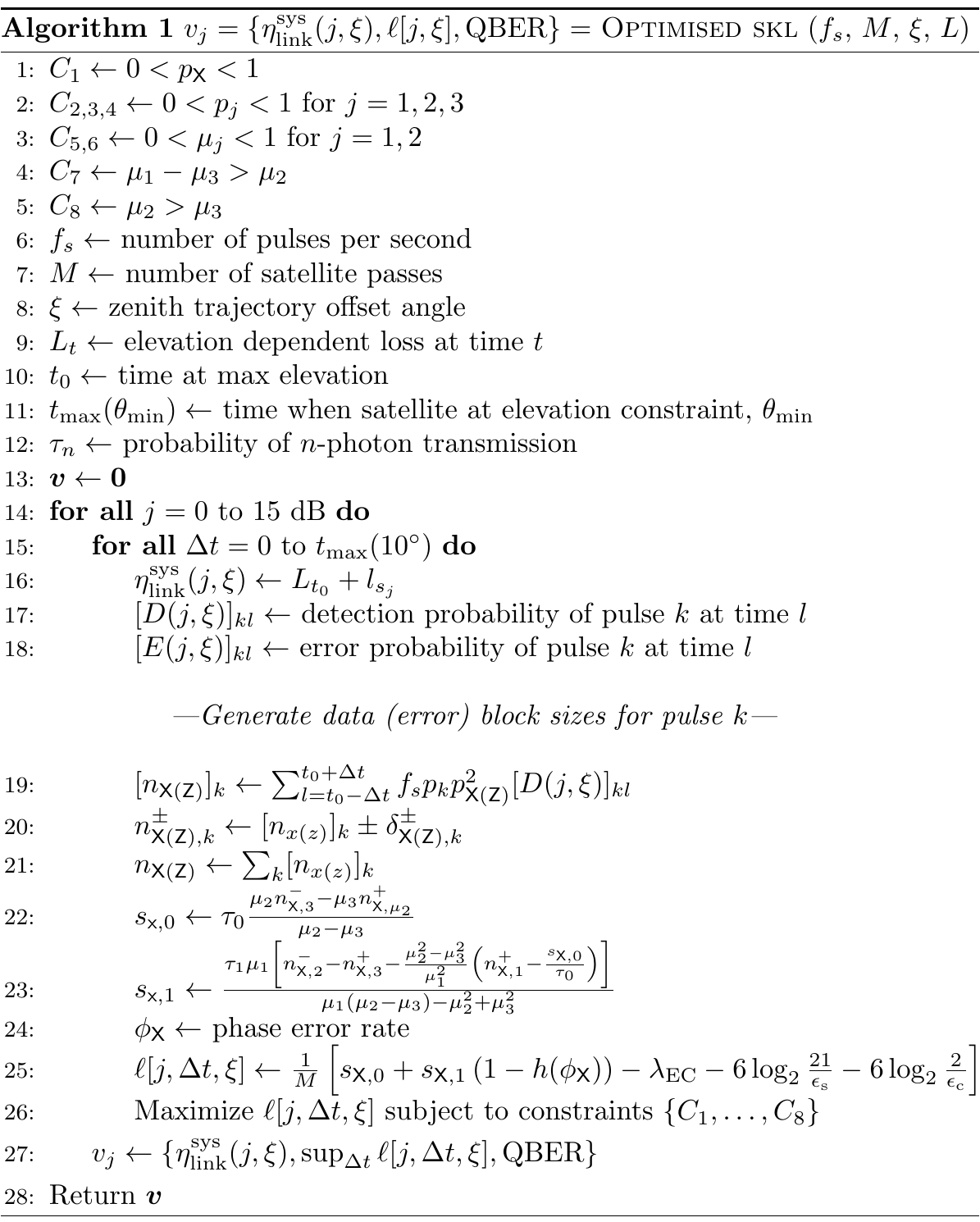}
    \caption{Pseudocode for the optimised finite key length for a pulse repetition rate $f_s$, $M$ satellite passes in the plane offset by angle $\xi$ from the zenith plane, and time dependent losses $L$. An elevation constraint of $\theta_\text{min}=10^\circ$ imposes realistic local constraints of establishing an optical link between the OGS and satellite. The key length is optimised over all protocol parameters and a transmission time window $\Delta t$ over which the raw key is acquired. The correction terms \smash{$\delta^{\pm}_{\mathsf{X}(\mathsf{Z}), k}$} for bases $\mathsf{X}(\mathsf{Z})$ and the intensities $\mu_k$ are determined from the multiplicative Chernoff bound that account for finite statistics~\cite{Yin2020_SR}.}
    \label{fig:pseudo_code_1}
\end{figure}

Due to limited transmission times, satellite-based quantum communications are strongly affected by finite statistics. To model different SatQKD systems, we improve the analysis in Ref.~\cite{Lim2014_PRA} with recent developments in modelling statistical fluctuations arising from finite statistics. This improvement leads to a more robust SKL and is imprinted through the finite statistic correction terms $\delta^{\pm}_{\mathsf{X(Z)},k}$, which we define using the inverse multiplicative Chernoff bound~\cite{Zhang2017_PRA,Yin2020_SR}. Specifically, let $Y$ denote a sum of $M$ independent Bernoulli samples, which need not be identical. Denote $y^{\infty}$ as the expectation value of $Y$, with $y$ the observed value for $Y$ from a single experimental run. The magnitude of difference between the observed and expected values depends on the statistics available. To quantify this deviation, we determine the probability that $y \le y^{\infty}+\delta^{+}_{Y}$ is less than a fixed positive constant $\varepsilon>0$, and the probability that $y \ge y^{\infty}-\delta^{-}_{Y}$ is less that $\varepsilon$. This is achieved through setting
\begin{align}
\delta^{+}_{Y}=\beta+\sqrt{2\beta y +\beta^2}, \quad
\delta^{-}_{Y}=\frac{\beta}{2}+\sqrt{2\beta y +\frac{\beta^2}{4}},
\end{align}
where $\beta=\ln(1/\varepsilon)$~\cite{Yin2020_SR}. Hence, we define the following finite sample size data block size
\begin{align}
\begin{split}
n^{\pm}_{\mathsf{X(Z)},k}&=\frac{e^k}{p_k}\left[n_{\mathsf{X(Z)},k}\pm \delta^{\pm}_{n_{\mathsf{X(Z)},k}} \right],\\
m^{\pm}_{\mathsf{X(Z)},k}&=\frac{e^k}{p_k}\left[m_{\mathsf{X(Z),k}}\pm \delta^{\pm}_{m_{\mathsf{X(Z)},k}} \right],
\end{split}
\end{align}
for the number of events and errors respectively in the $\mathsf{X}(\mathsf{Z})$ basis. From this, we define  the vacuum and single photon yields, and the phase error rate of single-photon events using Ref.~\cite{Lim2014_PRA} (see also pseudocode~\ref{fig:pseudo_code_1}) that determines the finite key length.

An important step in any QKD protocol is error correction. This necessitates classical communication, of $\lambda_\text{EC}$ bits, which are assumed known to Eve. This must be taken account of in the privacy amplification stage. While $\lambda_\text{EC}$ is known for practical implementations of the protocol, it must be estimated for the finite key optimisation. An overly conservative estimate would yield no key in a region of the parameter space where one should be viable. Conversely, an overly optimistic value for $\lambda_\text{EC}$ leads to spurious results. This highlights the importance of choosing a good estimate. It is standard to model the channel as a bit-flip channel \cite{Lim2014_PRA,Tomamichel2017_QIP}. This leads to $\lambda_\text{EC}=f_\text{EC} n_\mathsf{X} h_2(Q)$, where $Q$ is the QBER and $f_\text{EC}$ the reconciliation factor. The value for $f_\text{EC}$ is chosen slightly above unity, e.g. 1.16. The use of a constant reconciliation factor is a simple way of accounting for inefficiency in the error correction protocol.  This approach is normally sufficient when determining the optimal secret key length. However, for satellite QKD, one operates with high losses that are at the limit of where one can extract a key.  As such, it is beneficial to use a more refined estimate of $\lambda_\text{EC}$.

A better estimate for $\lambda_\text{EC}$ is given in Ref.~\cite{Tomamichel2017_QIP}.  In this approach, the correction to $n_{\mathsf{X}} h(Q)$ depends on the data block size. In particular,
\begin{align}
\begin{split}
    \lambda_\text{EC} = &n_{\mathsf{X}} h(Q) + n_{\mathsf{X}} (1 - Q)\log\left[\frac{(1 - Q)}{Q}\right]\\
    &- 
    \left(F^{-1}( \epsilon_{c};n_{\mathsf{X}},1-Q,) - 1\right) \log\left[\frac{(1 - Q)}{Q}\right]\\
    &- \frac{1}{2} \log(n_{\mathsf{X}}) - \log(1/\epsilon_{c}),
\end{split}
\label{eq:lambdaec}
\end{align}
where $n_{\mathsf{X}}$ is the data block size, $Q$ is the QBER and $F^{-1}$ is the inverse of the cumulative distribution function of the binomial distribution. We utilise this definition to estimate the information leaked during error correction in the finite key regime. 

These post-processing terms define the attainable finite key, subject to rigorous statistical analyses. The key length is a function of the basis encoding probability, $p_\mathsf{X}$, the source intensities and their probabilities, $\{\mu_j, p_j\}$ for $j\in \{1,2,3\}$, and the transmission time window, $\Delta t$, used to construct block data for a satellite pass. Without loss of generality, we set the second decoy state intensity as the vacuum $\mu_3=0$. For a defined SatQKD system, we generate an optimised finite key length by optimising over the parameter space of the six variables: $\{p_\mathsf{X}, \mu_1, \mu_2, p_1, p_2, \Delta t\}$. A baseline for system performance used in this work is detailed in Table~\ref{tab:loss_error_parameters} in the main text. This procedure can be generalised for any satellite trajectory. Fig.~\ref{fig:pseudo_code_1} illustrates a pseudocode of our numerical optimiser.

%


\subsection{Asymptotic key length per pass}
\label{sec:asymptotic}

\noindent
In this section, we provide an operational definition for the asymptotic key length per pass and explain how it can be determined by adapting the finite key optimiser. Data is combined from multiple satellite passes. The optimisation over the protocol parameters is then performed on the combined results and a secret key is extracted. For the following, we assume that each satellite pass has the same trajectory. With minimal modification, this can be extended to varying satellite trajectories for each pass.

Denoting $\ell_M$ as the SKL attained from $M$ satellite passes, we saw in Fig.~\ref{fig:SKL_multiple_passes} that $\ell_M \neq M \ell_1$. We also saw that the improvement to the SKL decreased with increasing number of satellite passes. This leads to a natural question: \emph{what is the largest attainable SKL per pass?} This is given by the asymptotic secret key length $\ell_{\infty}=\lim_{M\rightarrow\infty} \ell_{M}/M$.  The key length, $\ell_{M}$, is found using equation (\ref{eqn:skl_lim_result}), and the quantity $\ell_{\infty}$ is determined by examining the asymptotic scaling of $\ell_{M}/M$. 

The estimate for the vacuum counts per pass is~\cite{Lim2014_PRA},
\begin{align}
    \frac{s_{\mathsf{X,0}}}{M}=\frac{\tau_0}{\mu_2-\mu_3}\frac{\mu_2 \Gamma_3 \left(n_{\mathsf{X,3}}-\delta^{-}_{\mathsf{X},3}\right)-\mu_3 \Gamma_2\left(n_{\mathsf{X},2}+\delta^{+}_{\mathsf{X},2}\right)}{M},
\end{align}
where $n_{\mathsf{X},k}$ is the number of sifted counts in the $\mathsf{X}$-basis, from pulses of intensity $k$, $\tau_0$ is averaged probability that a vacuum state is transmitted by the laser,  $\Gamma_k=\exp(\mu_k)/p_k$ and $\delta^{\pm}_{\mathsf{X},k}$ are correction terms that account for the finite statistics. In Ref.~\cite{Lim2014_PRA}, $\delta^{\pm}_{\mathsf{X},k}$ is derived from Hoeffding's inequality. A higher SKL is attained by instead deriving these correction terms from the multiplicative Chernoff bound~\cite{Yin2020_SR}. The asymptotic scaling of these correction terms are $\smash{\mathcal{O}(\sqrt{n_{\mathsf{X}}})}$. This scaling is independent of whether the Hoeffding and Chernoff bounds are used. This scaling implies that the scaling with the number of satellite passes is $\mathcal{O}(\sqrt{M})$. Hence, $\delta^{\pm}_{\mathsf{X},k}/M$ scales $\smash{\mathcal{O}(1/\sqrt{M})}$ and thus tends to zero as $M\rightarrow \infty$. The finite statistics correction terms thus go to zero, as expected.

Since each satellite pass is assumed to have the same orbit, the total number of counts accumulated $n_{\mathsf{X},k}$ is equal to $M$ times the corresponding number of counts for a single pass, $n^{(1)}_{\mathsf{X},k}$. From this, we obtain 
\begin{align}
    \lim_{M\rightarrow\infty} \frac{s_{\mathsf{X,0}}}{M}=\frac{\tau_0}{\mu_2-\mu_3}\left(\mu_2 \Gamma_3 n^{(1)}_{\mathsf{X,3}}-\mu_3 \Gamma_2 n^{(1)}_{\mathsf{X},2}\right)=s^{\infty}_{\mathsf{X},0},
\end{align}
where $s^{\infty}_{\mathsf{X},0}$ is the asymptotic estimate of the vacuum counts for a single pass, which we define formally in the next paragraph.
By following a similar process for each term in $\ell_{M}/M$, we obtain
\begin{align}
    \ell_{\infty}=\Big\lfloor s^{\infty}_{\mathsf{X},0}+s^{\infty}_{\mathsf{X},1}\left(1-h(\phi^{\infty}_{\mathsf{X}})\right)-\lambda^{\infty}_\text{EC}\Big\rfloor,
\end{align}
where \smash{$\phi^{\infty}_{\mathsf{X}}=\nu^{\infty} _{\mathsf{Z},1}/s^{\infty}_{\mathsf{Z},1}$} is the phase error rate, $s^{\infty}_{\mathsf{X},1}$, $s^{\infty}_{\mathsf{Z},1}$ and $\nu^{\infty}_{\mathsf{Z},1}$ are the single pass asymptotic estimates for: the single photon counts in the $\mathsf{X}$ basis, the $\mathsf{Z}$ basis, and the single photon errors in the $\mathsf{Z}$ basis respectively. 

The asymptotic quantities: \smash{$\nu^{\infty} _{\mathsf{Z},1}$, $s^{\infty}_{\mathsf{X},0}$ and $s^{\infty}_{\mathsf{X(Z)},1}$}, correspond to averaging the single pass quantities over infinitely many passes.  More formally, let $u\in\{\phi_{\mathsf{X}}, \nu_{\mathsf{Z},1}, s_{\mathsf{X(Z)},0},s_{\mathsf{X(Z)},1},\lambda_\text{EC} \}$, and let $u^{(M)}$ denote the quantity estimated using the full data from $M$ passes.  The asymptotic quantity is then defined as 
\begin{align}
    u^{\infty}=\lim_{M\rightarrow\infty} \frac{u^{(M)}}{M}.    
\end{align}

A refined estimate of $\lambda_\text{EC}$ and its upper bound on the asymptotic behaviour is provided in Ref.~\cite{Tomamichel2017_QIP}. From this, we determine that \smash{$\lambda^{\infty}_\text{EC}=n^{(1)}_{\mathsf{X}} h(Q)$}, where $Q$ is the QBER for a single pass. When running the finite key optimiser for the asymptotic key length per pass, we define $\lambda^{\infty}_\text{EC}=1.16 n_{\mathsf{X}}^{(1)} h(Q)$, which accounts for inefficient error correction even in the asymptotic limit. 

Rather than looking at the key length per pass, it is also common to consider the key rate, i.e. the number of secret key bits per transmitted pulse.  Let $N$ be the total number of pulses transmitted by the satellite during a single pass.  The key rate for $M$ passes is just $\text{SKR}_{M}=\ell_{M}/(MN)$.  In the limit of infinitely  many passes, the asymptotic key rate is given by $\text{SKR}_{\infty}=\ell_{\infty}/N$.

Notice that our analysis of the asymptotic secret key rate, $\text{SKR}_{\infty}$ differs from the route often taken in the literature. Specifically, the asymptotic key rate can be determined as a function of different elevation angles. The data is then combined according to Eq.~\eqref{eq:sklasymptotic}.  While such an approach is possible if we extract keys for each angle of elevation, it is not appropriate for the current analysis, where we group all the data for a pass and then extract a key from the combined data.

\onecolumngrid
\setcounter{section}{0}
\renewcommand{\thesection}{\Roman{section}}

\setcounter{subsection}{0}

\setcounter{figure}{0}
\renewcommand{\thefigure}{E\arabic{figure}}

\newpage

\section*{EXTENDED DATA}

\begin{figure}[h!]
    \centering
    \includegraphics[width=\linewidth]{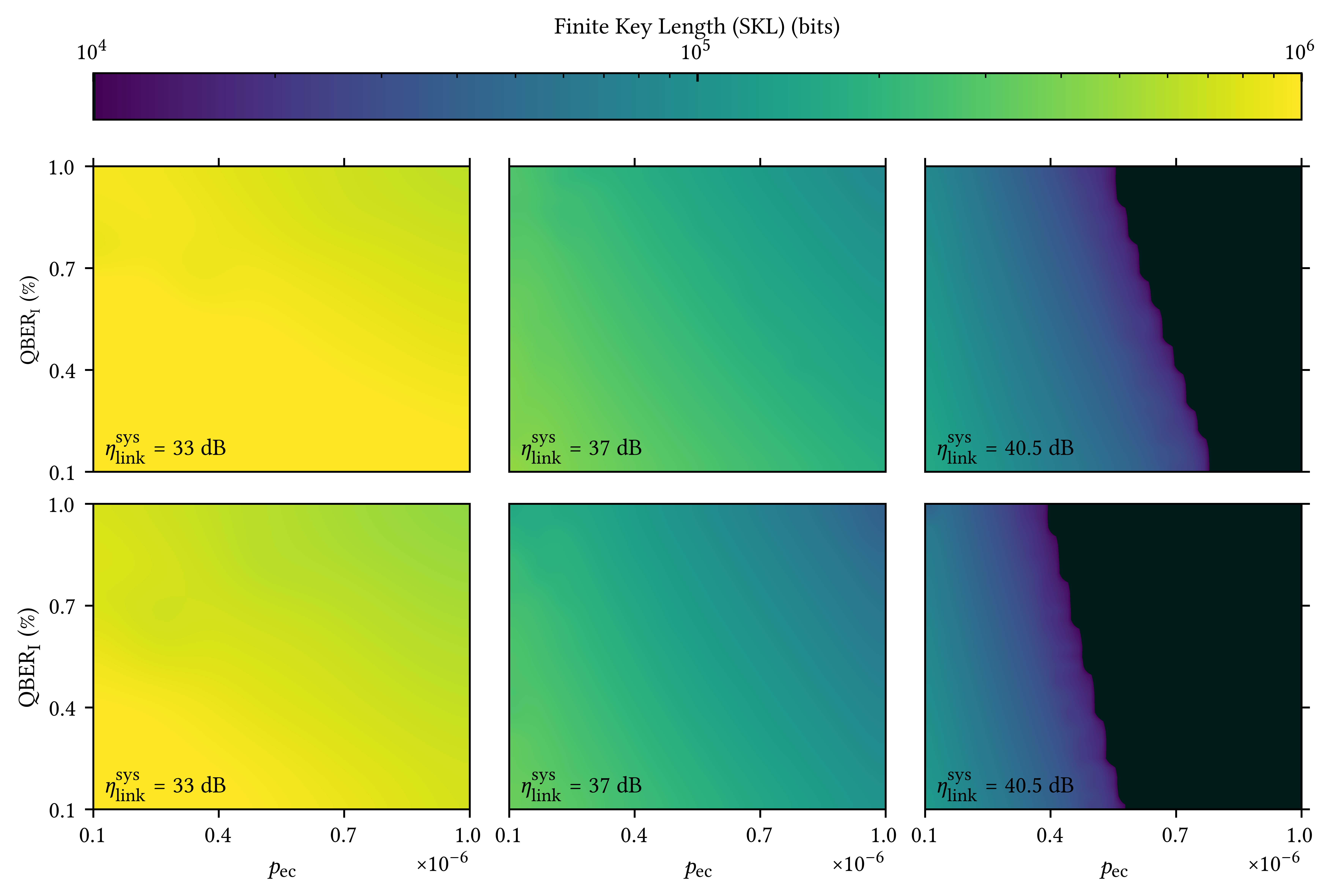}
    \caption{SKL vs $p_\text{ec}$ and $\text{QBER}_\text{I}$. We consider different $\sle$ values and overpasses geometries. The first row corresponds to an ideal zenith overpass, and the second to an example of a non-ideal overpass with maximum elevation $\theta_\text{max} = 60^\circ$. For good system link efficiencies, the SKL shows little change across the parameter space. In contrast, for more modest $\sle$ values, the SKL exhibits greater sensitivity to changes in the system parameters, specifically $p_\text{ec}$. At $\sle=40.5$ dB the optimised SKL rapidly reduces to zero for both overpass geometries. The black region indicates the parameter space where no finite key is attainable.}
    \label{fig:contour_plot}
\end{figure}%
%


%
\begin{figure}[h!]
    \centering
    \includegraphics[width=0.55\textwidth]{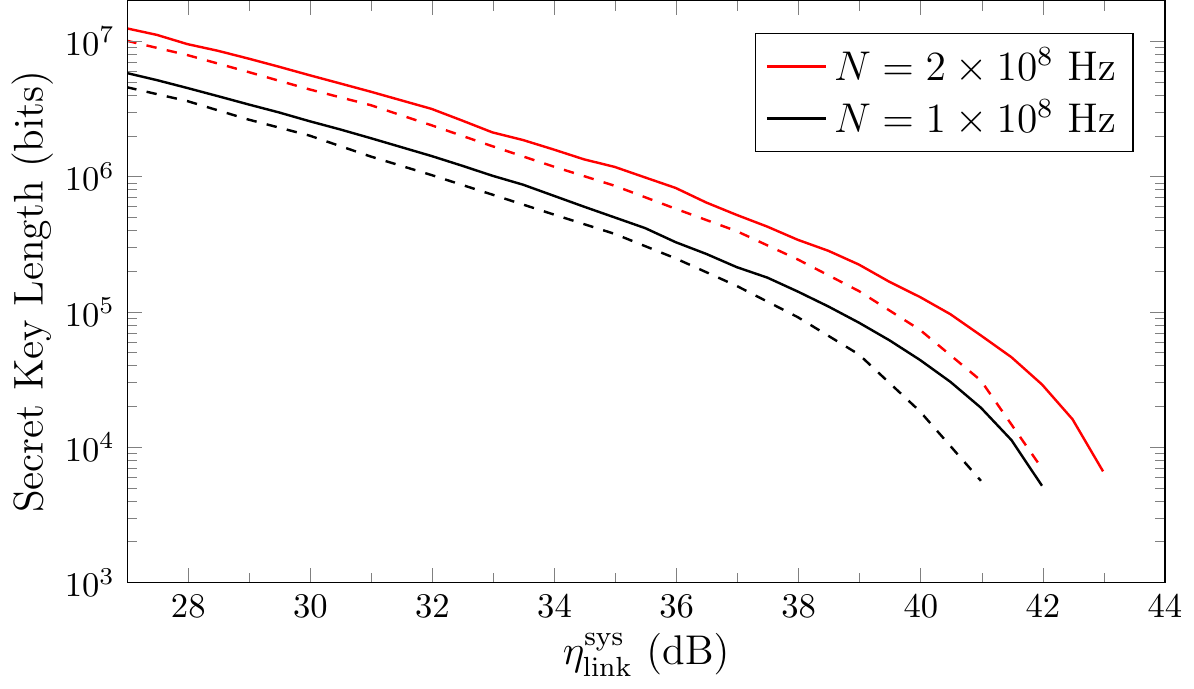}
    \caption{SKL versus system link efficiency for different source repetition rates. $\text{QBER}_\text{I}=0.5\%$ and $p_\text{ec}=5\times 10^{-7}$ per pulse. Solid lines correspond to a zenith overpass and dashed lines to a non-ideal overpass with $\theta_\text{max}=60^\circ$. Doubling the source repetition rate increases the total number of pulses transmitted depending on the duration of the overpass, which increases the SKL. A system with $f_s=100$ MHz at system link efficiency $\sle$ provides a similar SKL as a system operating with $f_s=200$ MHz but with a 3 dB worse system link efficiency. For the same $\sle$, doubling the source rate can lead to more than double the SKL. Not only is the sifted key length twice as long but the parameter uncertainties are reduced, leading to smaller finite-key effects. There is also a relative improvement due to better error correction efficiency and reduced overhead of the composable security parameters. }
    \label{fig:npulse}
\end{figure}%
%


%
\begin{figure}[h!]
    \centering
    \includegraphics[width=0.55\columnwidth]{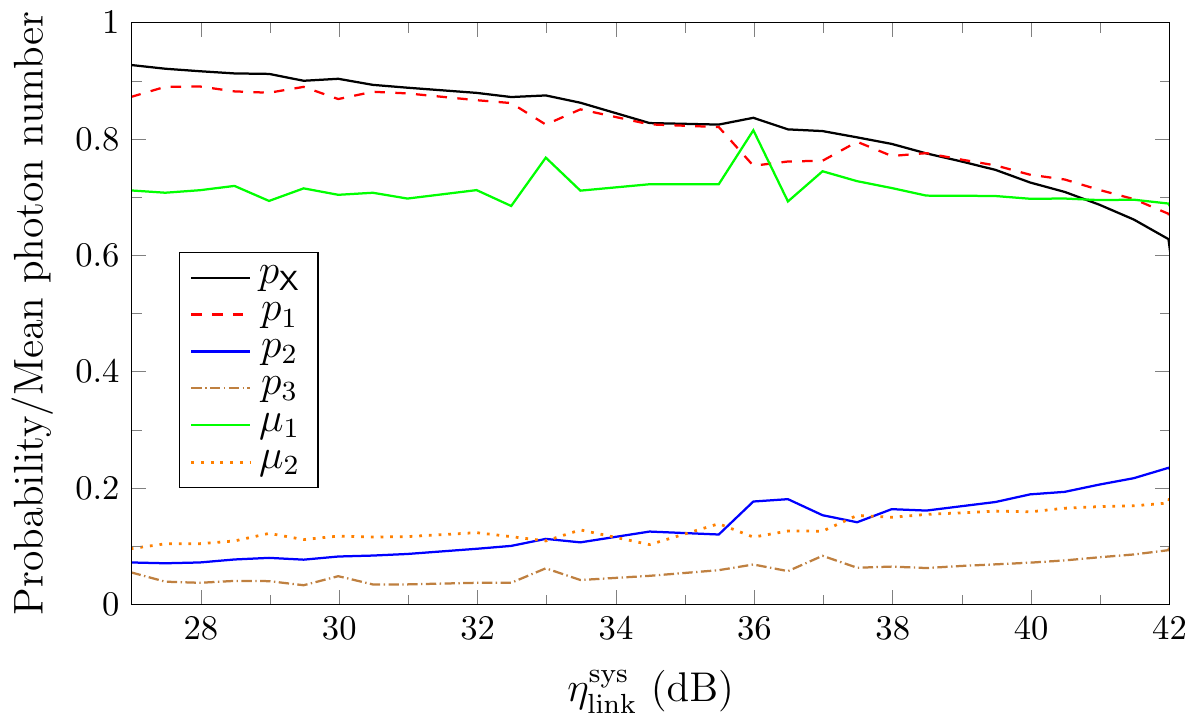}
    \caption{Optimised decoy-state protocol parameters with system link efficiency. Single zenith overpass with $\smash{p_\text{ec} = 5 \times 10^{-7}}$ and $\text{QBER}_\text{I}$ = 0.5\%. Protocol parameters are fixed for the duration of a single pass so that all received signals can be processed together. As $\sle$ increases, we see that $p_{\mathsf{X}}$ decreases. This is because more signal pulses must be transmitted in the $\mathsf{Z}$ basis to overcome the larger quantum channel losses.}
    \label{fig:optimisation_parameters}
\end{figure}%
\begin{figure}[h!]
    \centering
    \includegraphics[width=0.55\textwidth]{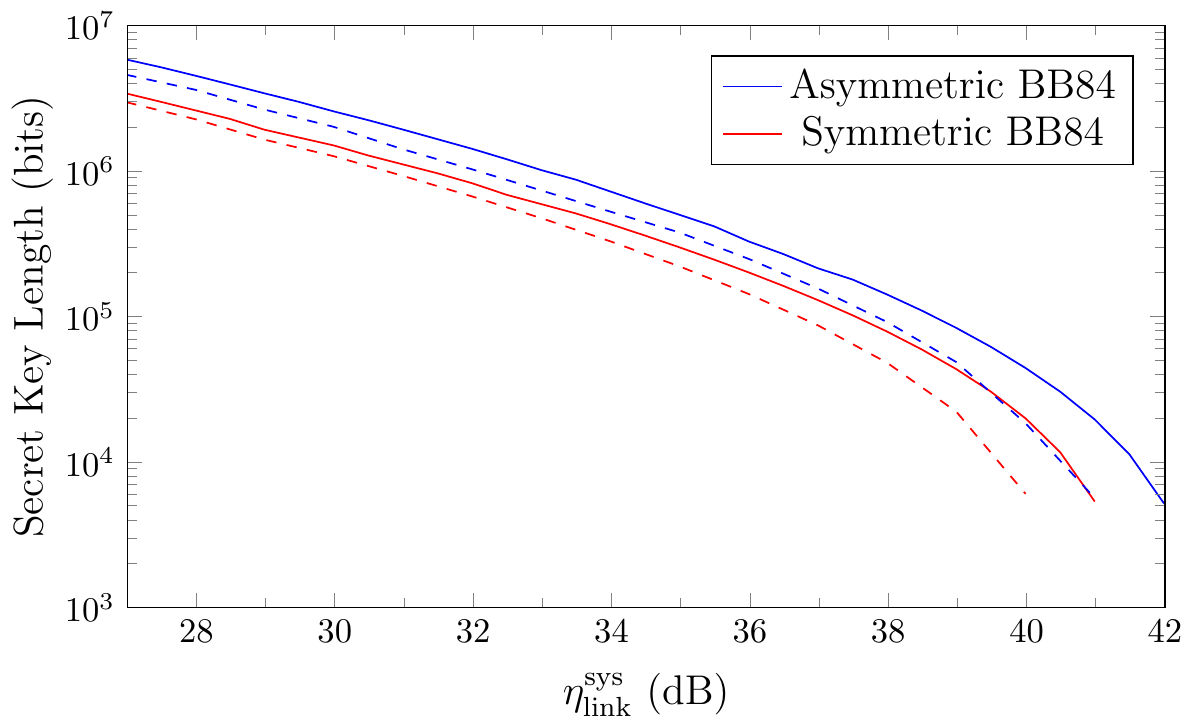}
    \caption{SKL vs \smash{$\sle$} for a-BB84 and s-BB84 protocols. System parameters are \smash{$p_\text{ec}=5\times 10^{-7}$}, and  $\text{QBER}_\text{I} = 0.5\%$. Solid lines correspond to a zenith pass and dashed lines to a non-ideal overpass with maximum elevation of $\theta_\text{max}=60^\circ$.}
    \label{fig:asym_sym_loss}
\end{figure}%

\end{document}